\documentclass[prb,reprint,twocolumn,amsmath,amssymb]{revtex4-1}

\newcommand{\del}{\nabla}

\newcommand{\bvPsi}{\boldsymbol{\varPsi}}

\newcommand{\bx}{\boldsymbol{\textbf{x}}}

\newcommand{\bH}{\boldsymbol{\textbf{H}}}
\newcommand{\bI}{\boldsymbol{\textbf{I}}}

\newcommand{\bM}{\boldsymbol{\textbf{M}}}

\newcommand{\bN}{\boldsymbol{N}}

\newcommand{\bP}{\boldsymbol{\textbf{P}}}
\newcommand{\bR}{\boldsymbol{\textbf{R}}}
\newcommand{\bS}{\boldsymbol{\textbf{S}}}

\newcommand{\bX}{\boldsymbol{X}}
\newcommand{\bY}{\boldsymbol{Y}}

\newcommand{\bPhi}{\boldsymbol{\Phi}}

\newcommand{\bGam}{\boldsymbol{\Gamma}}

\newcommand{\dx}{\,d\bx}

\newcommand{\bbH}{\bar{\boldsymbol{\textbf{H}}}}

\newcommand{\op}[1]{\mathcal{#1}}
\newcommand{\order}{\mathcal{O}}
\renewcommand{\arraystretch}{1.2}
\newcommand{\ket}[1]{\left| #1 \right>} 
\newcommand{\bra}[1]{\left< #1 \right|} 
\newcommand{\braket}[2]{\left< #1 \vphantom{#2} \right | \left. #2 \vphantom{#1} \right>} 
\newcommand{\matrixel}[3]{\left< #1 \vphantom{#2#3} \right| #2 \left| #3 \vphantom{#1#2} \right>} 
\usepackage{graphicx}
\usepackage{graphics}
\usepackage{multirow}
\usepackage{fancyhdr, ifpdf}
\usepackage{amsxtra}
\usepackage{stmaryrd}
\usepackage{mathrsfs}
\usepackage{psfrag}
\usepackage{subfigure}
\usepackage{epsfig}
\usepackage{rotating}
\usepackage{setspace}
\usepackage{float}
\usepackage{amsfonts}
\usepackage{amsbsy}
\usepackage{amscd}
\usepackage{amsthm}
\usepackage{color}
\usepackage{tabularx}
\usepackage{caption}
\usepackage{sidecap}
\usepackage{dcolumn}
\usepackage[ruled]{algorithm2e}
\usepackage{algorithmic}

\definecolor{hellgruen}{rgb}{0.2,0.7,0.2}

\begin{document}
\preprint{}
 \title{Spectrum-splitting approach for Fermi-operator expansion in all-electron Kohn-Sham DFT calculations}
\author{Phani Motamarri}
\author{Vikram Gavini}
\affiliation{Department of Mechanical Engineering, University of Michigan, Ann Arbor, MI 48109, USA}
\author{Kaushik Bhattacharya}
\author{Michael Ortiz}
\affiliation{Division of Engineering and Applied Science, California Institute of Technology, Pasadena, CA 91125, USA}
\begin{abstract}
We present a spectrum-splitting approach to conduct all-electron Kohn-Sham density functional theory (DFT) calculations by employing Fermi-operator expansion of the Kohn-Sham Hamiltonian. The proposed approach splits the subspace containing the occupied eigenspace into a core-subspace, spanned by the core eigenfunctions, and its complement, the valence-subspace, and thereby enables an efficient computation of the Fermi-operator expansion by reducing the expansion to the valence-subspace projected Kohn-Sham Hamiltonian. The key ideas used in our approach are: (i) employ Chebyshev filtering to compute a subspace containing the occupied states followed by a localization procedure to generate non-orthogonal localized functions spanning the Chebyshev-filtered subspace; (ii) compute the Kohn-Sham Hamiltonian projected onto the valence-subspace; (iii) employ Fermi-operator expansion in terms of the valence-subspace projected Hamiltonian to compute the density matrix, electron-density and band energy. We demonstrate the accuracy and performance of the method on benchmark materials systems involving silicon nano-clusters up to 1330 electrons, a single gold atom and a six-atom gold nano-cluster. The benchmark studies on silicon nano-clusters revealed a staggering five-fold reduction in the Fermi-operator expansion polynomial degree by using the spectrum-splitting approach for accuracies in the ground-state energies of $\sim10^{-4} Ha/atom$ with respect to reference calculations. Further, numerical investigations on gold suggest that spectrum splitting is indispensable to achieve meaningful accuracies, while employing Fermi-operator expansion.

\end{abstract}
\maketitle
\section{Introduction}\label{sec:intro}
In the present day, quantum-mechanically informed calculations on ground-state materials properties are readily possible by means of electronic-structure calculations -via- the Kohn-Sham density functional theory (DFT) framework~\cite{kohn64,kohn65}. Over the past few decades, DFT has been instrumental in providing significant insights into materials properties across a range of materials systems. The approach to DFT is based on the key result by Hohenberg and Kohn, who in their seminal work~\cite{kohn64} showed that the ground-state properties of a materials system can be described by a functional of electron-density, which, to date, remains unknown. This challenge has however been addressed by Kohn and Sham ~\cite{kohn65} in an approximate sense by reducing the many-body problem of interacting electrons into an equivalent problem of non-interacting electrons in an effective mean field that is governed by the electron-density. This effective single-electron formulation accounts for the quantum-mechanical interaction between electrons by means of an unknown exchange-correlation term which is modeled in practice, and the widely used models have been successful in predicting a range of properties across various materials systems.

The self-consistent field (SCF) approach to solving the Kohn-Sham DFT problem involves, in any given SCF iteration, discretizing the Kohn-Sham eigenvalue problem using an appropriate basis set followed by the diagonalization of the discrete Kohn-Sham Hamiltonian to obtain the eigenvalues and orthonormal eigenvectors, which are then used to compute the electron-density. The computational complexity of solving the Kohn-Sham DFT problem using conventional approaches typically scales as $\order{(M\,N^2)}$ where $M$ denotes the number of basis functions and $N$ denotes the system size (number of atoms or number of electrons).  As $M \propto N$, the computational cost, which is cubic scaling with system size, becomes prohibitively expensive. This has led to numerous efforts focused on the development of methods~\cite{goedecker99,bow2012} aimed at reducing the computational complexity of DFT calculations, which include density matrix minimization methods~\cite{vanderbilt93,haynes2006}, divide and conquer methods~\cite{yang92,Ozaki2006,barrault2007}, Fermi-operator expansion techniques~\cite{goedecker93,baer97,linlin2009,linlin2013a}, Fermi-operator approximation method~\cite{phanish2013a,ponga2016}, Fermi-operator projection method~\cite{sankey94,stephan98}, orbital minimization approach~\cite{mauri93,kim95,gao2009} and subspace projection type methods~\cite{carlos,motam2014,Tucker2016}. Either the locality of the representation of the wavefunctions, or the exponential decay of density matrix in real-space is generally exploited in these methods. It has been demonstrated that these methods work well for insulating systems exhibiting linear-scaling with system size. However, the computational complexity of these approaches can deviate significantly from linear-scaling, in practice, for metallic systems. Further, some of the developed techniques~\cite{vanderbilt93,sankey94,stephan98,carlos} assume the existence of a band-gap, thus restricting these techniques solely to insulating systems. 

One reduced-order scaling technique that is equally applicable to both insulating and metallic systems at finite temperatures is the Fermi-operator expansion method~\cite{goedecker93,baer97,goedecker99}, which computes the finite-temperature density matrix through a Chebyshev polynomial approximation of the Fermi-Dirac function of the Kohn-Sham Hamiltonian. The width of the eigenspectrum ($\Delta E$) of the discretized Hamiltonian and the smearing parameter ($\sigma = k_B\,T$) in the Fermi-Dirac function determine the accuracy of such an expansion. Numerous recent efforts~\cite{ozaki2007,ceriotti2008,linlin2009,ozaki2010,linlin2013a} have focused on developing methods that aim to reduce the number of terms used in the expansion to approximate the Fermi-Dirac function. 

While the Fermi-operator expansion method has been widely employed in pseudopotential calculations, there are major challenges in using this technique for all-electron DFT calculations. 
One of the challenges is that the discrete Hamiltonians, employing real-space basis sets like finite-elements~\cite{motam2013} or wavelets~\cite{BigDFT}, for all-electron DFT calculations have very large spectral widths of $\order(10^6)$ due to the refined discretizations that are needed to resolve the rapidly oscillating wavefunctions around the nuclei. The Fermi-operator expansion method is inaccurate and impractical for such large spectral widths. This issue has recently been addressed in Motamarri et al.~\cite{motam2014} by employing the Fermi-operator expansion on a subspace projected Hamiltonian whose spectral width is commensurate with that of the occupied eigenspectrum.   A second challenge in the case of all-electron calculations arises from the large width of the occupied spectrum, especially in materials systems involving atoms with large atomic numbers, which can again render the Fermi-operator expansion inaccurate. We remark that the magnitude of the smallest algebraic eigenvalue grows as $\order(Z^2)$ where $Z$ denotes the largest atomic number among all the atoms comprising the materials system. 

In this paper, we overcome these challenges by proposing a spectrum-splitting approach for all-electron Kohn-Sham DFT calculations in which the subspace containing the occupied eigenspace is split into a core-subspace, and its complement, referred to as the valence-subspace. Subsequently, the Fermi-operator expansion is efficiently calculated by reducing the expansion to the valence-subspace projected Hamiltonian. In this paper, the proposed method is discussed in the context of spectral finite-element discretization. However, the method presented is general enough to be applicable to any real-space numerical discretization employed in the solution of the all-electron Kohn-Sham DFT problem. The main ideas constituting our approach are: (i) employ a Chebyshev filter to compute a subspace containing the occupied eigenspace, followed by a localization procedure to generate non-orthogonal localized functions spanning the Chebyshev-filtered subspace; (ii) compute the projection of the Hamiltonian onto the valence-subspace; (iii) employ the Fermi-operator expansion in terms of the valence-subspace projected Hamiltonian to compute the density matrix, electron-density and band energy.

We begin by presenting an abstract mathematical framework where the projection of the Hamiltonian onto the valence-subspace is derived in terms of the projection matrix corresponding to the core-subspace and the Chebyshev-filtered subspace projected Hamiltonian. These projections are expressed in a non-orthogonal localized basis spanning the Chebyshev-filtered subspace, which is instrumental in realizing a reduced-order scaling numerical implementation of DFT~\cite{carlos,motam2014} by taking advantage of the sparsity of the resulting matrices. We then derive the expressions for the density matrix and the constraint on the number of valence electrons in terms of the valence-subspace projected Hamiltonian, which are subsequently used to develop the spectrum-splitting approach for finite-element discretized all-electron Kohn-Sham DFT calculations. To this end, we first represent the Kohn-Sham Hamiltonian and the corresponding wavefunctions in the L{\"o}wdin orthonormalized finite-element basis constructed using spectral finite-elements in conjunction with Gauss-Lobatto-Legendre quadrature rules. The self-consistent field (SCF) iteration begins with the action of a Chebyshev filter on a given initial subspace spanned by localized single-atom wavefunctions to compute an approximation to the occupied eigenspace of the discretized Kohn-Sham Hamiltonian. A localization procedure is then employed to construct non-orthogonal localized functions spanning the Chebyshev-filtered subspace. The Kohn-Sham Hamiltonian is then projected onto the Chebyshev-filtered subspace expressed in the localized basis. We subsequently employ a second Chebyshev filtering procedure on the subspace projected Hamiltonian to compute an approximation to the core-subspace, followed by a localization procedure to construct non-orthogonal localized functions spanning the core-subspace. These localized functions are in turn employed to compute the projection matrix corresponding to the core-subspace, which is then used to evaluate the projected Hamiltonian corresponding to the valence-subspace. The Fermi-operator expansion of the discrete Kohn-Sham Hamiltonian, which is used to compute the finite-temperature density matrix and the electron-density, is reduced to an expansion on the valence-subspace projected discrete Hamiltonian whose spectral width is $\order(1)$.

We investigate the accuracy and performance of the proposed method on representative benchmark materials systems involving silicon nano-clusters up to 1330 electrons, gold atom and a six-atom gold nano-cluster. In the case of silicon nano-clusters, the proposed spectrum-splitting approach resulted in a staggering five-fold reduction in the Fermi-operator expansion polynomial degree for desired accuracies of $\sim10^{-4} Ha/atom$ in the ground-state energies. The utility of the spectrum-splitting approach is even more evident in the benchmark calculations on gold, which has a high atomic number (Z=79). Our results indicate that, by employing the spectrum-splitting approach, accuracies of $\sim 10^{-4} Ha/atom$ can be achieved using a Fermi-operator expansion polynomial degree of around $1000$. However, if the Fermi-operator expansion is employed directly on the Chebyshev-filtered subspace projected Hamiltonian, even a polynomial degree of 2000 resulted in errors of $\order(1\,Ha)$ in ground-state energies per atom. This suggests that spectrum-splitting is indispensable while employing Fermi-operator expansion in all-electron DFT calculations on materials systems containing atoms with high atomic numbers. 

The remainder of the paper is organized as follows. Section II describes the real-space formulation of the all-electron Kohn-Sham DFT problem. Section III presents the mathematical framework for spectrum-splitting, and derives the relevant expressions that will be used subsequently. Section IV describes the various steps involved in the spectrum-splitting algorithm within the framework of spectral finite-element discretization. Section V discusses the numerical studies on benchmark examples demonstrating the accuracy and performance of the approach. We finally conclude with a summary and outlook in Section VI.

\section{All-electron Kohn-Sham density functional theory}\label{sec:formulation}
Consider a materials system comprising of $N_a$ nuclei and $N_e$ electrons. In density functional theory, the variational problem of computing ground-state properties of a given materials system is equivalent to solving the following non-linear Kohn-Sham eigenvalue problem~\cite{kohn65}:
\begin{equation}\label{contEigen}
\left(-\frac{1}{2} \del^{2} + V_{\text{eff}}(\rho,\bR)\right) {\psi}_{i} = \epsilon_{i} {\psi}_{i},\;\;\;\;\; i = 1,2,\cdots
\end{equation}
where $\epsilon_{i}$ and $\psi_{i}$ denote the eigenvalues and corresponding eigenfunctions of the Hamiltonian, respectively. We denote by $\bR = \{\bR_1,\,\bR_2,\,\cdots \bR_{N_a}\}$ the collection of all nuclear positions in the materials system. Though the proposed ideas in this work can be easily generalized to periodic or semi-periodic materials systems and spin-dependent Hamiltonians~\cite{rmartin}, for the sake of simplicity, we discuss the formulation in the context of non-periodic setting restricting to spin-independent Hamiltonians. 

The electron-density at any spatial point $\bx$, in terms of the eigenfunctions, is given by
\begin{equation}
\rho(\bx) = 2\sum_{i}f(\epsilon_i,\mu)|\psi_i(\bx)|^2 \,\,,
\end{equation}
where $f(\epsilon_i,\mu)$ is the orbital occupancy function, whose range lies in the interval $\left[0,1\right]$, and $\mu$ represents the Fermi-energy. In density functional theory calculations, it is fairly common to represent $f$ by the Fermi-Dirac distribution~\cite{VASP,goedecker99} 
\begin{equation}\label{fermidirac1}
f(\epsilon,\mu) = \frac{1}{1 + \exp\left(\frac{\epsilon - \mu}{\sigma} \right)}\,,
\end{equation}
where $\sigma = k_{B} T$ is a smearing parameter with $k_B$ and $T$ denoting the Boltzmann constant and the finite temperature. We note that as $T\searrow0$, the Fermi-Dirac distribution tends to the Heaviside function. The Fermi-energy $\mu$ is computed from the constraint on the total number of electrons in the system ($N_e$) given by
\begin{equation}\label{cons}
\int \rho(\bx) \dx = 2\sum_i f(\epsilon_i,\mu) = N_e\,.
\end{equation}
The effective single-electron potential in the Kohn-Sham Hamiltonian in equation~\eqref{contEigen}, $V_{\text{eff}}(\rho,\bR)$, is given by the sum of exchange-correlation potential ($V_{\text{xc}}(\rho)$) that accounts for the quantum-mechanical interactions and the classical electrostatic potentials corresponding to electron-density ($V_{\text{H}}(\rho)$) and nuclear charges ($ V_{\text{ext}}(\bR)$):
\begin{equation}\label{veff}
 V_{\text{eff}}(\rho,\bR) = V_{\text{xc}}(\rho) +V_{\text{H}}(\rho) +V_{\text{ext}}(\bR)\,.
\end{equation}
In the present work, the exchange-correlation interactions are treated using the local-density approximation (LDA)~\cite{rmartin}, and in particular we employ the Slater exchange and Perdew-Zunger~\cite{alder,perdew} form of correlation functional. In equation~\eqref{veff}, the classical electrostatic potential corresponding to the electron-density distribution (Hartree potential), $V_{\text{H}}(\rho)$, is given by
\begin{equation}
V_\text{H}(\rho) (\bx) = \int \frac{\rho(\bx')}{|\bx - \bx'|} \dx'\,.
\end{equation}
Further, electrostatic potential corresponding to the nuclear charges, $ V_{\text{ext}}(\bx,\bR)$, is given by
\begin{equation}\label{extpot}
V_{\text{ext}}(\bx,\bR)  = -\sum_{I=1}^{N_a} \frac{Z_I}{|\bx - \bR_I|} \,,
\end{equation}
with $Z_{I}$ denoting the atomic number of the $I^{th}$ nucleus in the given materials system. We note that the computation of electrostatic potentials $V_H$ and $V_{\text{ext}}$ are extended in real-space. However, these quantities can be efficiently computed by taking recourse to the solution of the Poisson problem~\cite{arias2000,surya2010,motam2013}, since the kernel corresponding to these extended interactions is the Green's function of the Laplace operator. Finally, for given positions of nuclei, the system of equations corresponding to the Kohn-Sham eigenvalue problem are:
\begin{subequations}\label{ksproblem}
\begin{gather}
\left(-\frac{1}{2} \del^{2} +  V_{\text{xc}}(\rho) + V_{\text{H}}(\rho) + V_{\text{ext}}(\bR) \right){\psi}_{i} = \epsilon_{i} {\psi}_{i},\\
2\sum_{i}f(\epsilon_i,\mu) = N_e,\\
\rho(\bx) = 2\sum_{i}f(\epsilon_i,\mu)|\psi_{i}(\bx)|^2.
\end{gather}
\end{subequations}
We note that the system of equations in~\eqref{ksproblem} represents a nonlinear eigenvalue problem, which has to be solved self-consistently. Upon self-consistently solving~\eqref{ksproblem}, the ground-state energy of the system is given by
\begin{align}
E_{\text{tot}} =\,& E_{\text{band}} +E_{\text{xc}} (\rho) - \int V_{\text{xc}}(\rho)\, \rho \dx  \notag\\
&  -\frac{1}{2}\int \rho V_{\text{H}}(\rho) \dx + E_{\text{ZZ}}\,,
\end{align}
where $E_{\text{xc}}$ denotes the exchange-correlation energy functional corresponding to $V_{\text{xc}}$, $E_{\text{band}}$ denotes the band energy, which is given by
\begin{equation}
E_{\text{band}} = 2 \sum_{i} f(\epsilon_i,\mu) \epsilon_i\,,
\end{equation}
and $E_{\text{ZZ}}$ denotes the nuclear-nuclear repulsive energy, which is given by
\begin{equation}
E_{\text{ZZ}} = \frac{1}{2} \sum_{I,J=1 \atop I \neq J}^{N_a} \frac{Z_I Z_{J}}{|\bR_{I} - \bR_{J}|}\,.
\end{equation}

\section{Mathematical Formulation}\label{sec:math}
In this section, we first derive the expression for the projection of the Hamiltonian operator onto the valence-subspace of the Kohn-Sham Hamiltonian. Subsequently, we derive the expressions for the computation of electron-density and constraint on the number of valence electrons in terms of the valence-subspace projected Hamiltonian. 

Let $\mathbb{H}$ denote the infinite-dimensional Hilbert space of functions equipped with the inner product  $\left<.|.\right>$ and a norm $\parallel. \parallel$ derived from the inner product. If the Hermitian operator $\op H$ defined on $\mathbb{H}$ denotes the Kohn-Sham Hamiltonian of interest at a step of the self-consistent iteration, then the projection of $\op H$ onto an $M$-dimensional subspace $\mathbb{V}_h^{M}\subset \mathbb{H}$ is given by $\op {\tilde{H}} \equiv \op P^q \op H \op P^q: \mathbb{H} \rightarrow \mathbb{V}_h^{M}$ where $\op P^q:\mathbb{H} \rightarrow \mathbb{V}_h^{M}$ denotes the projection operator onto $\mathbb{V}_h^{M}$ and is given by $\op P^q = \sum_{i=1}^{M}\ket{q_i}\bra{q_i}$. Here, $\{\ket {q_i}\}_{1\leq i \leq M}$ denote an orthonormal basis for the subspace $\mathbb{V}_h^{M}$ with a real-space representation $\braket{\bx}{q_i}=q_i(\bx)$ for $i=1,\dots,M$. The L{\"o}wdin orthonormalized finite-element basis, which is employed in the present work, can be one choice for the basis functions.
\vspace{-0.3in}
\subsection{Operator splitting in eigen-subspace}
\vspace{-0.1in}
Let $\mathbb{V}^{N} \subset \mathbb{V}_h^{M}$ denote the eigen-subspace of $\op{\tilde{H}}$ that includes all the occupied states as well as the unoccupied states around the Fermi energy. A good approximation to this eigen-subspace can be computed, for instance, using a Chebyshev filtering approach ~\cite{saad2006}. Let $\{ \ket {\phi_{\alpha}}\}_{1\leq \alpha \leq N}$ denote a non-orthogonal basis spanning the eigen-subspace. The generality of using a non-orthogonal basis to represent the eigen-subspace is motivated from our eventual objective of using non-orthogonal localized basis functions, computed using a localization procedure~\cite{motam2014}, to realize reduced-order scaling computational complexity for the solution of the Kohn-Sham problem. Let the projection of $\op {\tilde{H}}$ onto $\mathbb{V}^{N}$ be denoted by $\op H^{\phi} \equiv\op P^\phi \op {\tilde{H}} \op P^\phi: \mathbb{V}_h^{M} \rightarrow \mathbb{V}^{N}$, where $\op P^\phi:\mathbb{V}_h^{M} \rightarrow \mathbb{V}^{N}$ is the projection operator onto $\mathbb{V}^{N}$ and is given by $\op P^\phi = \sum_{\alpha,\beta=1}^{N}\ket{\phi_\alpha}S^{-1}_{\alpha\beta}\bra{\phi_\beta}$. Here, $S_{\alpha \beta} = \braket {\phi_\alpha} {\phi_\beta}$ are the matrix elements of the overlap matrix denoted by $\bS$. 

We now consider the spectral decomposition of $\op H^{\phi}$
\begin{equation}\label{projHam}
\op H^{\phi} = \sum_{i=1}^{N} \epsilon^{\phi}_i \ket {\psi_i^{\phi}} \bra {\psi_i^{\phi}}
\end{equation}
with $\epsilon^{\phi}_i$ and $\ket {\psi_i^{\phi}}$ denoting the eigenvalues and the corresponding eigenvectors of $\op H^{\phi}$, respectively. Splitting the projected Hamiltonian $\op H^{\phi}$ in ~\eqref{projHam} into a core part and its complement, referred to as the valence part, we have
\begin{align}
\op H^{\phi} &= \sum_{i=1}^{N_{\text{core}}}\epsilon_i^{\phi} \ket{\psi_i^{\phi}} \bra{\psi_i^{\phi}} + \sum_{i=N_{\text{core}}+1}^{N} \epsilon_i^{\phi} \ket{\psi_i^{\phi}} \bra{\psi_i^{\phi}}\nonumber \\
 &= \op H^{\phi,\text{core}} + \op H^{\phi,\text{val}}
\end{align}
In the above, $\op H^{\phi,\text{core}}$ and $\op H^{\phi,\text{val}}$ denote the projection of $\op H^{\phi}$ onto the core- and valence-subspaces denoted by $\mathbb{V}^{N_{\text{core}}}$ and $\mathbb{V}^{N_{\text{val}}}$, respectively. Here $N_{\text{core}}$ denotes the number of core states, while $N_{\text{val}}= N -N_{\text{core}}$. We now derive an expression for $\op H^{\phi,\text{val}}$, and, to do so, we denote the non-orthogonal basis spanning $\mathbb{V}^{N_{\text{core}}}$ to be $\{\ket {\phi_{\alpha}^{\text{core}}}\}_{1\leq \alpha \leq N_{core}}$. The projection operator onto $\mathbb{V}^{N_{\text{core}}}$, $\op P^{\phi,\text{core}}:\mathbb{V}^{N} \rightarrow \mathbb{V}^{N_{\text{core}}}$, can be expressed as
\begin{equation}
\op P^{\phi,\text{core}} = \sum_{i=1}^{N_{\text{core}}} \ket{\psi_i^{\phi}} \bra{\psi_i^{\phi}} = \sum_{\alpha,\beta=1}^{N_{\text{core}}}\ket{\phi_\alpha^{\text{core}}}S^{\text{core}^{-1}}_{\alpha\beta}\bra{\phi_\beta^{\text{core}}}\,,
\end{equation}
where $S_{\alpha\beta}^{\text{core}} = \braket {\phi^{\text{core}}_\alpha} {\phi^{\text{core}}_\beta}$ denote the matrix elements of the core overlap matrix denoted by $\bS^{\text{core}}$. Further, the projection operator onto the space $\mathbb{V}^{N_{\text{val}}}$ is given by $(\op I -\op P^{\phi,\text{core}}):\mathbb{V}^{N} \rightarrow \mathbb{V}^{N_{\text{val}}}$ with $\op I$ denoting the identity operator. Thus, the projection of $\op H^{\phi}$ onto $\mathbb{V}^{N_{\text{val}}}$, $\op H^{\phi,\text{val}} :\mathbb{V}^{N} \rightarrow \mathbb{V}^{N_{\text{val}}}$, is given by
\begin{equation}
\op H^{\phi,\text{val}} = (\op I -\op P^{\phi,\text{core}}) \op H^{\phi} (\op I -\op P^{\phi,\text{core}}).
\end{equation}
Let us now consider the single particle density operator ($\varGamma^{\phi}$) corresponding to $\op H^{\phi}$, which can be expressed as
\begin{equation}\label{denop}
\varGamma^{\phi} = \sum_{i=1}^{N} f(\epsilon^{\phi}_i) \ket {\psi_i^{\phi}} \bra {\psi_i^{\phi}} = f(\op H^{\phi})\,,
\end{equation}
where the second equality follows from the spectral decomposition of the Hermitian operator $\op H^{\phi}$ in ~\eqref{projHam}. Splitting the density operator $\varGamma^{\phi}$ into core and valence parts, we have
\begin{align}\label{densitymatsplit}
\varGamma^{\phi} &= \sum_{i=1}^{N_{\text{core}}}f(\epsilon_i^{\phi}) \ket{\psi_i^{\phi}} \bra{\psi_i^{\phi}} + \sum_{i=N_{\text{core}}+1}^{N} f(\epsilon_i^{\phi}) \ket{\psi_i^{\phi}} \bra{\psi_i^{\phi}}\nonumber \\
 &=  \varGamma^{\phi,\text{core}} + \varGamma^{\phi,\text{val}}\,.
\end{align}
Using the fact that the core-states have unit occupancy, i.e. $f(\epsilon_i^{\phi})= 1$ for $i=1 \cdots N_{\text{core}}$, equation~\eqref{densitymatsplit} can be reduced to
\begin{equation}\label{vargamma}
\varGamma^{\phi} = f(\op H^{\phi}) = \op P^{\phi,\text{core}} + f(\op H^{\phi,\text{val}})\,.
\end{equation}
\vspace{-0.2in}
\subsection{Matrix representations in non-orthogonal basis}
In this subsection, we derive the matrix expressions for the operators $\op H^{\phi,\text{val}}$, $\op P^{\phi,\text{core}}$ and $\varGamma^{\phi,\text{val}}$ expressed in the non-orthogonal basis $\{\ket {\phi_{\alpha}}\}_{1\leq \alpha \leq N}$, which are used in the spectrum-splitting algorithm described in the next section. To this end, we denote the matrices corresponding to $\op H^{\phi}$ and $\op H^{\phi,\text{val}}$ in $\{\ket {\phi_{\alpha}}\}_{1\leq \alpha \leq N}$ basis as $\bH^{\phi}$ and $\bH^{\phi,\text{val}}$, respectively. The matrix elements of $\bH^{\phi}$ are given by
\begin{equation}
\text{H}^{\phi}_{\alpha \beta}  = \sum_{\gamma = 1} ^{N}  S^{-1}_{\alpha \gamma} \matrixel {\phi_\gamma} {\op H^{\phi}}  {\phi_\beta} \,.\nonumber 
\end{equation}
The matrix elements of $\bH^{\phi,\text{val}}$ are given by
\begin{align}\label{compRalpbet}
&\text{H}^{\phi,\text{val}}_{\alpha \beta}  = \sum_{\gamma = 1} ^{N}  S^{-1}_{\alpha \gamma} \matrixel {\phi_\gamma} {\op H^{\phi,\text{val}}}  {\phi_\beta} \nonumber \\
& = \sum_{\gamma = 1}^{N}  S^{-1}_{\alpha \gamma} \matrixel {\phi_\gamma} {(\op I - \op {P^{\phi,\text{core}}}) \op {H^{\phi}} (\op I - \op {P^{\phi,\text{core}}})} {\phi_\beta}\,. \nonumber  
\end{align}
Using the relation $\op {H^{\phi}} = \op P^\phi \op {H^{\phi}} \op P^\phi$ in the above equation, we have
\begin{align}
&\text{H}^{\phi,\text{val}}_{\alpha \beta} = \nonumber \\
&\sum_{\gamma=1}^{N} S^{-1}_{\alpha \gamma} \matrixel {\phi_\gamma} {(\op I - \op P^{\phi,\text{core}})\op P^{\phi} \op H^{\phi} \op P^{\phi} (\op I - \op P^{\phi,\text{core}})} {\phi_\beta} \,.\nonumber
\end{align}
We now use the definition of $\op P^\phi$ to simplify the above equation as
\begin{equation}
\text{H}^{\phi,\text{val}}_{\alpha \beta} = \sum_{\nu,\eta=1}^{N} (\delta_{\alpha \nu} - \text{P}^{\phi,\text{core}}_{\alpha \nu}) \text{H}^{\phi}_{\nu \eta}(\delta_{\eta \beta}-\text{P}^{\phi,\text{core}}_{\eta \beta})\,,
\end{equation}
where $\text{P}^{\phi,\text{core}}_{\alpha \nu}$ denotes the matrix element of $\bP^{\phi,\text{core}}$, which is the matrix corresponding to operator $\op P^{\phi,\text{core}}$ expressed in $\{\ket {\phi_{\alpha}}\}_{1\leq \alpha \leq N}$ basis. Further, $\delta_{\alpha \nu}$ denotes the Kronecker delta. Thus, using matrix notation, we have
\begin{equation}\label{Hphival}
\bH^{\phi,\text{val}} = (\bI - \bP^{\phi,\text{core}})\bH^{\phi}(\bI - \bP^{\phi,\text{core}}) \,.
\end{equation}

We now derive an expression for the matrix element of $\bP^{\phi,\text{core}}$ as follows:
\begin{align}
\text{P}^{\phi,\text{core}}_{\alpha \beta} &= \sum_{\gamma = 1}^{N}S^{-1}_{\alpha \gamma}\matrixel {\phi_{\gamma}} {\op P^{\phi,\text{core}}} {\phi_\beta}\nonumber \\
& = \sum_{\gamma = 1}^{N}S^{-1}_{\alpha \gamma} \matrixel {\phi_{\gamma}} {\op P^q\: \op P^{\phi,\text{core}}\: \op P^q} {\phi_\beta}\nonumber
\end{align}
Substituting the expressions $\op P^q = \sum_{i=1}^{M}\ket{q_i}\bra{q_i}$ and $\op P^{\phi,\text{core}} = \sum_{\alpha,\beta=1}^{N_{\text{core}}}\ket{\phi_\alpha^{\text{core}}}S^{\text{core}^{-1}}_{\alpha\beta}\bra{\phi_\beta^{\text{core}}}$ into the above equation, we have
\begin{equation}
\text{P}^{\phi,\text{core}}_{\alpha \beta} = \sum_{\gamma=1}^{N}\sum_{\eta,\nu=1}^{N_{\text{core}}}\sum_{i,j=1}^{M}S^{-1}_{\alpha \gamma} \Phi^{*}_{i\gamma}\Phi^{\text{core}}_{i\eta}S^{\text{core}^{-1}}_{\eta\nu}\Phi^{\text{core}^*}_{j\nu}\Phi_{j\beta}\,,
\end{equation}
where $\Phi_{j\beta} = \braket{q_j}{\phi_\beta}$ and $\Phi^{*}_{i\gamma}$ denotes the complex conjugate of $\Phi_{i\gamma}$, and the above equation can be conveniently recast in terms of matrices as follows 
\begin{align}
\bP^{\phi,\text{core}} &= \bS^{-1}\bPhi^{\dagger}\bPhi^{\text{core}}\bS^{\text{core}^{-1}}\bPhi^{\text{core}^{\dagger}}\bPhi \nonumber\\
&=\widetilde{\bPhi}^{\text{core}}\bS^{\text{core}^{-1}}\widetilde{\bPhi}^{\text{core}^{\dagger}}\bS \,, \label{projOp}
\end{align}
where $\widetilde{\bPhi}^{\text{core}}=\bS^{-1}\bPhi^{\dagger}\bPhi^{\text{core}}$ represents the matrix with column vectors comprising of the components of $\{\ket{\phi^{\text{core}}_{\alpha}}\}_{1\leq\alpha\leq N_{\text{core}}}$ expressed in $\{\ket {\phi_{\alpha}}\}_{1\leq \alpha \leq N}$ basis. Here, $\bPhi$ denotes the matrix whose column vectors are the components of $\{\ket {\phi_{\alpha}}\}_{1\leq \alpha \leq N}$ in $\{\ket {q_i}\}_{1\leq i \leq M}$ basis, and $\bPhi^{\dagger}$ denotes the conjugate transpose of the matrix $\bPhi$.  Furthermore, $\bPhi^{\text{core}}$ denotes the matrix whose column vectors are the components of $\{\ket{\phi^{\text{core}}_{\alpha}}\}_{1\leq\alpha\leq N_{\text{core}}}$ expressed in $\{\ket {q_i}\}_{1\leq i \leq M}$ basis. $\bS^{\text{core}}$, which denotes the core overlap matrix, can be expressed in terms of $\widetilde{\bPhi}^{\text{core}}$ as
\begin{equation}
\bS^{\text{core}} =  \widetilde{\bPhi}^{\text{core}^{\dagger}}\bS \widetilde{\bPhi}^{\text{core}}\,.
\end{equation}

Next, we derive the matrix expression for $\varGamma^{\phi,\text{val}}$ in the non-orthogonal basis. We denote the matrix representation of $\varGamma^{\phi,\text{val}}$ in $\{\ket{\phi_{\alpha}}\}_{1 \leq \alpha \leq N}$ basis and $\{\ket{q_{i}}\}_{1 \leq i \leq M}$ by $\bGam^{\phi,\text{val}}$ and $\bGam^{q,\text{val}}$, respectively. We seek to express $\bGam^{\phi,\text{val}}$ in terms of $\bH^{\phi,\text{val}}$. We have 
\begin{align}
&\Gamma^{\phi,\text{val}}_{\alpha \beta} = \sum_{\gamma = 1}^{N}S^{-1}_{\alpha \gamma}\matrixel {\phi_{\gamma}} {\varGamma^{\phi,\text{val}}} {\phi_\beta}\nonumber\\
&= \sum_{\gamma=1}^{N} S^{-1}_{\alpha \gamma} \matrixel {\phi_{\gamma}} {\op P^q\: \varGamma^{\phi,\text{val}}\: \op P^q} {\phi_\beta}\nonumber\\
&=\sum_{\gamma=1}^{N}\sum_{i,j=1}^{M} S^{-1}_{\alpha \gamma}\braket {\phi_{\gamma}} {q_i}\matrixel {q_i} {\varGamma^{\phi,\text{val}}} {q_j}\braket{q_j}{\phi_\beta}\nonumber\\
&= \sum_{\gamma=1}^{N}\sum_{i,j=1}^{M} S^{-1}_{\alpha \gamma} \Phi^{*}_{i\gamma}\Gamma^{q,\text{val}}_{ij}\Phi_{j\beta} \,.
\end{align}
We note that $\bGam^{q,\text{val}}=f(\bH^{q,\text{val}})$, where $\bH^{q,\text{val}}$ is the matrix representation of $\op H^{\phi,\text{val}}$ in $\{\ket {q_i}\}_{1 \leq i \leq M}$. Hence, in matrix notation, we have,
\begin{equation}\label{gammval}
\bGam^{\phi,\text{val}}= \bPhi^{+}f(\bH^{q,\text{val}})\bPhi\,,
\end{equation}
where $\bPhi^{+} = \bS^{-1}\bPhi^{\dagger}$ denotes the Moore-Penrose pseudoinverse of $\bPhi$ matrix. We now seek a relation between $\bH^{q,\text{val}}$ and $\bH^{\phi,\text{val}}$ for which we use the relations $\op {H^{\phi,\text{val}}} = \op P^\phi \op {H^{\phi,\text{val}}} \op P^\phi$ and $\op P^\phi = \sum_{\mu,\delta=1}^{N}\ket{\phi_\mu}S^{-1}_{\mu\delta}\bra{\phi_\delta}$ to express $\bH^{q,\text{val}}$ as
\begin{align}
\text{H}^{q,\text{val}}_{ij} &= \matrixel {q_i} {\op {H^{\phi,\text{val}}}} {q_j}\nonumber \\
&= \matrixel {q_i} {\op P^\phi \op {H^{\phi,\text{val}}} \op P^\phi} {q_j}\nonumber\\
&= \sum_{\alpha,\delta = 1}^{N}\Phi_{i\alpha}\text{H}^{\phi,\text{val}}_{\alpha \mu}S^{-1}_{\mu \delta}\Phi^{*}_{j\delta}\,.
\end{align}
Using matrix notation, the above equation can be written as
\begin{equation}
\bH^{q,\text{val}} = \bPhi \bH^{\phi,\text{val}} \bPhi^{+}\,.
\end{equation}
We note that the function $f(\epsilon)$ represents a Fermi-Dirac distribution, which is an analytic function, and, hence, $f(\bH^{q,\text{val}})$ admits a power series representation given by
\begin{align}
\bGam^{q,\text{val}} &= f(\bH^{q,\text{val}}) = \sum_{k=0}^{\infty} a_k (\bH^{q,\text{val}})^k = \sum_{k=0}^{\infty} a_k (\bPhi  \bH^{\phi,\text{val}} \bPhi^{+})^k\nonumber\\
&= \sum_{k=0}^{\infty} a_k \bPhi (\bH^{\phi,\text{val}})^{k} \bPhi^{+} = \bPhi \Bigl(\sum_{k=0}^{\infty} a_k (\bH^{\phi,\text{val}})^{k}\Bigr) \bPhi^{+}\nonumber\\
&= \bPhi f(\bH^{\phi,\text{val}}) \bPhi^{+}\,.
\end{align}
Substituting the above relation in equation~\eqref{gammval}, we arrive at
\begin{equation}\label{gammavalmatrix}
\bGam^{\phi,\text{val}} = f(\bH^{\phi,\text{val}})\,.
\end{equation}
Hence, using equations ~\eqref{vargamma}, ~\eqref{projOp} and ~\eqref{gammavalmatrix} we have
\begin{equation}\label{gammphi}
\bGam^{\phi} = f(\bH^{\phi}) = \widetilde{\bPhi}^{\text{core}}\bS^{\text{core}^{-1}}\widetilde{\bPhi}^{\text{core}^{\dagger}}\bS + f(\bH^{\phi,\text{val}})\,.
\end{equation}
Finally, we derive the constraint on the number of valence electrons, denoted by $N_{e,\text{val}}$, using equation~\eqref{gammphi}, by recalling that the constraint on the total number of electrons is given by 
\begin{align*}
&2\,\text{tr}(f(\bH^{\phi})) = N_e \nonumber \\
\implies &2\,\text{tr}(\bP^{\phi,\text{core}}) + 2\,\text{tr}(f(\bH^{\phi,\text{val}})) = N_e\nonumber \,,\\
\implies &2\,\text{tr}(f(\bH^{\phi,\text{val}}) = N_e - 2\,N_{\text{core}}\,.
\end{align*}
Hence, the constraint on the number of valence electrons $N_{e,\text{val}}$ is given by
\begin{equation}\label{valconst}
2\,\text{tr}(f(\bH^{\phi,\text{val}})) = N_{e,\text{val}} = N_e - 2\,N_{\text{core}}\,.
\end{equation}

\section{Spectrum-splitting algorithm}\label{sec:split}
In this section, we present the various steps involved in the spectrum-splitting approach for conducting all-electron Kohn-Sham DFT calculations within the finite-element framework. The subspace projection technique proposed in Motamarri et.al~\cite{motam2014} is used as the starting point in developing the proposed approach, which involves: (i) computing the discretized Hamiltonian in L{\"o}wdin orthonormalized finite-element basis; (ii) computing the relevant eigen-subspace using Chebyshev filtering, followed by evaluating the non-orthogonal localized basis spanning the Chebyshev filtered basis; (iii) computing the projected Hamiltonian in the non-orthogonal basis to evaluate the quantities of interest such as the density-matrix, electron-density and band energy using Fermi-operator expansion. The key idea in this work is to reformulate the Fermi-operator expansion of the Kohn-Sham Hamiltonian in terms of the valence-subspace projected Hamiltonian. The spectral width of the valence-subspace projected Hamiltonian, which is $\order(1Ha)$, is significantly smaller than the spectral widths of both the discrete Kohn-Sham Hamiltonian in the finite-element basis as well as the Chebyshev-filtered subspace projected Hamiltonian. Further, the spectral width of the valence-subspace projected Hamiltonian is relatively independent of the basis set discretization and the materials system. As demonstrated subsequently, this significantly improves the accuracy of Fermi-operator expansion type techniques, which have been developed to realize reduced-order scaling for the Kohn-Sham problem. A brief summary of the ideas proposed in Motamarri et.al~\cite{motam2014} is first discussed in this section, followed by the description of the proposed spectrum-splitting approach.
\subsection{Projection of the Kohn-Sham Hamiltonian in the L{\"o}wdin orthonormalized finite-element basis}
Let $\mathbb{V}_h^{M}$ be the $M$ dimensional subspace spanned by the finite-element basis, a piecewise polynomial basis generated from a finite-element discretization~\cite{Brenner-Scott} of characteristic mesh-size $h$. The various electronic fields namely the wavefunctions and the electrostatic potential expressed in the finite-element basis are given by
\begin{equation}\label{fem_phi}
\psi^{h}_{i}(\bx) = \sum_{j=1}^{M} N_j(\bx) \psi^{j}_{i}\,\,\,\,,\,\,\,\,\phi^{h}(\bx) = \sum_{j=1}^{M} N_j(\bx) \phi^{j}\,,
\end{equation}
where $N_{j}:1\leq j \leq M$ denotes the finite-element basis spanning $\mathbb{V}_h^{M}$. We now consider the L{\"o}wdin orthonormalized finite-element basis $q_j(\bx): {1 \leq j \leq M}$ spanning the finite-element space  $\mathbb{V}_h^{M}$ and we note the following relation between $q_j(\bx)$ and $N_j(\bx)$:
\begin{equation}\label{orthfem}
q_j(\bx) = \sum_{k=1}^{M} M_{jk}^{-1/2} N_k(\bx)\,,
\end{equation}
where $\bM$ denotes the overlap matrix associated with the basis functions $N_{j}:1\leq j \leq M$  with matrix elements $\text{M}_{jk} = \braket {N_j}{N_k}$. The discretization of the Kohn-Sham eigenvalue problem~\eqref{ksproblem} using this basis $q_j(\bx): {1\leq j \leq M}$ results in the following standard eigenvalue problem: 
\begin{equation}\label{ghep}
\bbH \bar{\bvPsi}_{i} = \epsilon^{h}_{i} \bar{\bvPsi}_{i}\,,
\end{equation}
where
\begin{equation}\label{discreteeigen}
\bbH = \bM^{-1/2}\bH\bM^{-1/2}
\end{equation}
with
\begin{gather}
\text{H}_{jk} = \frac{1}{2} \int \del N_{j} (\bx) .\del N_k(\bx) \dx \nonumber+\int V^{h}_{\text{eff}}(\bx) N_{j}(\bx) N_{k}(\bx) \dx \,, \\
\text{M}_{jk} = \int N_j(\bx) N_k(\bx) \dx \,. \notag
\end{gather}
Here $\bar{\bvPsi}_{i}$ is a vector containing the expansion coefficients of the discretized eigenfunction $\psi^{h}_i(\bx)$ expressed in L{\"o}wdin orthonormalized finite-element basis spanning the finite-element space. Furthermore, the matrix $\bM^{-1/2}$ can be evaluated with modest computational cost by using a spectral finite-element basis in conjunction with Gauss-Lobatto-Legendre (GLL) quadrature for the evaluation of integrals in the overlap matrix, which renders the overlap matrix diagonal~\cite{motam2013}.

\subsection{Chebyshev-filtered subspace iteration and localization}\label{sec:CFSILoc}
Chebyshev-filtered subspace iteration (ChFSI) technique~\cite{saad2006} is used to compute an approximation to the eigenspace of the discrete Kohn-Sham Hamiltonian, spanned by $N \geq N_e/2$ lowest eigenfunctions corresponding to the occupied states and a few unoccupied states around the Fermi energy. We refer to~\cite{motam2013,motam2014} for the details of its implementation in the context of finite-element discretization. The fast growth of Chebyshev polynomials in $(-\infty, -1)$ is exploited in the ChFSI technique to magnify the relevant eigenspectrum, and thereby providing an efficient approach for the solution of the Kohn-Sham eigenvalue problem. Further, a localized basis for the subspace $\mathbb{V}^{N}$ spanned by Chebyshev filtered vectors is computed by employing a localization technique as proposed in Garcia-Cervera et al.~\cite{carlos}. We refer to Motamarri et.al~\cite{motam2014} for the details of the numerical implementation involved in the computation of these localized functions. In the subsequent paragraphs, we denote by $\bPhi_{L}$ the matrix whose column vectors are the expansion coefficients of these compactly-supported non-orthogonal localized functions expressed in the L{\"o}wdin orthonormalized finite-element basis.

\subsection{Spectrum-splitting approach for computing electron-density}
We now discuss the steps involved in the spectrum-splitting approach involved in the computation of the electron-density. 
\paragraph{\textbf{Computation of projected Hamiltonian in the occupied subspace:}}
As discussed in section ~\ref{sec:math}, the projection of the Hamiltonian into the non-orthogonal localized basis $\bPhi_L$ is computed as 
\begin{equation}\label{projHamalpA}
\bH^{\phi} =  \bS^{-1} \bPhi^{T}_{L} \bbH \bPhi_{L}\,.
\end{equation}
where $\bS^{-1}$ denotes the inverse of the overlap matrix $\bS$ resulting from the non-orthogonal localized functions and is given by $\bS = \bPhi^{T}_L \bPhi_{L}$. 
\vspace{0.1in}
\paragraph{\textbf{Computation of the projection matrix corresponding to core states:}}
As discussed in section ~\ref{sec:math}, the relevant subspace of the Kohn-Sham Hamiltonian, which is of dimension $N$, can be split into two subspaces $\mathbb{V}^{N_{\text{core}}}$ and $\mathbb{V}^{N_{\text{val}}}$ spanned by the core- and valence-subspaces, respectively. Further, let $n^{\text{core}}_I$ be the number of core states associated with an isolated atom $I$ in the given material system, then the dimension of the subspace $\mathbb{V}^{N_{\text{core}}} \subset  \mathbb{V}^{N}$ is given by $N_{\text{core}} = \sum_{I=1}^{N_a} n^{\text{core}}_I$. We now seek to compute an approximation to the core-subspace, $\mathbb{V}^{N_{\text{core}}}$, spanned by the core eigenfunctions in order to evaluate the matrix corresponding to projection operator $\bP^{\phi,\text{core}}$. To this end, we employ Chebyshev filtering procedure to compute the core-subspace by constructing a filter from the projected Hamiltonian $\bH^{\phi}$. We start with an initial subspace denoted by the matrix $\bX^{\phi}$ of size $N \times N_{\text{core}}$, which is chosen to correspond to the core states from the localization procedure of section~\ref{sec:CFSILoc}. Subsequently, the projected Hamiltonian $\bH^{\phi}$ is scaled and shifted to construct $\widetilde{\bH}^{\phi}$ such that the valence spectrum of  $\bH^{\phi}$ is mapped to $[-1,1]$ and the core spectrum into $(-\infty,-1)$. Hence
\begin{equation}
\widetilde{\bH}^{\phi} = \frac{1}{g}(\bH^{\phi} - d\:\bI)\;\;\;\text{where}\;\;\;g = \frac{\epsilon_{\text{max}}^{\phi}-\epsilon^{\phi}}{2}\;\;\;d=\frac{\epsilon^{\phi}+\epsilon_{\text{max}}^{\phi}}{2}.
\end{equation}
where $\epsilon^{\phi}$ and $\epsilon_{\text{max}}^{\phi}$ denote the upper bounds of the core spectrum and the full spectrum of $\bH^{\phi}$, respectively. Estimate for $\epsilon_{\text{max}}^{\phi}$ can be computed using the Krylov-Schur method~\cite{kschur2001}. Further, $\epsilon^{\phi}$ is chosen to be lying in the energy-gap between $c_{\text{max}}$ and $v_{\text{min}}$, where $c_{\text{max}}$ denotes the maximum of the largest single-atom core eigenvalues associated with each atom $I$ in the given materials system, while $v_{\text{min}}$ denotes the minimum of the smallest single-atom valence eigenvalues. Finally, the filter is constructed using a Chebyshev polynomial of degree $m$, $T_m(x)$, and the action of the filter on $\bX^{\phi}$ is recursively computed as
\begin{equation}
\bY^{\phi} = T_m(\widetilde{\bH}^{\phi})\bX^{\phi} = \left[2\,\widetilde{\bH}^{\phi} T_{m-1}(\widetilde{\bH}^{\phi})  - T_{m-2}(\widetilde{\bH}^{\phi})\right]\bX^{\phi} \,.
\end{equation}
The Chebyshev-filtered vectors in $\bY^{\phi}$ can be expressed in terms of L{\"o}wdin orthonormalized basis by using transformation $\bY^{q} = \bPhi_L \bY^{\phi}$. 

Next, the localization procedure discussed in step B is employed to obtain non-orthogonal localized functions spanning the Chebyshev-filtered space $\bY^{q}$. Specifically, the localized functions are obtained by solving the minimization problem
\begin{equation}\label{loc}
\text{arg}\;\; \min_{\psi \in \mathbb{V}^{N_{\text{core}}}, ||\psi|| = 1} \int w(\bx) |\psi(\bx)|^2 \dx\,,
\end{equation}
where $w(\bx)  \geq  0$ is chosen to be a smooth weighting function  $|\bx - \bR_I|^{2}$ with the position of the nuclei $\bR_I$ denoting the localization center. We denote by $\bPhi^{\text{core}}_{L}$ the matrix whose column vectors are the expansion coefficients of these compactly supported non-orthogonal localized functions spanning the core-subspace expressed in the L{\"o}wdin orthonormalized basis. In order to compute the projection operator corresponding to core-subspace, it is computationally efficient to express the relevant matrices with respect to the basis (column) vectors of $\bPhi_L$. To this end, as discussed in section ~\ref{sec:math}, we express $\bPhi^{\text{core}}_{L}$ with respect to the basis $\bPhi_L$ by means of the following transformation
\begin{equation}
\widetilde{\bPhi}_L^{\text{core}}=\bS^{-1}\bPhi_L^{\dagger}\bPhi_L^{\text{core}}.
\end{equation}
Finally, the projection operator (cf. equation ~\eqref{projOp}) is computed as
\begin{equation}
\bP^{\phi,\text{core}} =\widetilde{\bPhi}_L^{\text{core}}\bS^{\text{core}^{-1}}\widetilde{\bPhi}_L^{\text{core}^{\dagger}}\bS \,,
\end{equation}
where $\bS^{\text{core}}$ in the above equation denotes the core overlap matrix which can be evaluated as
\begin{equation}
\bS^{\text{core}} =  \widetilde{\bPhi}_L^{\text{core}^{\dagger}}\bS \widetilde{\bPhi}_L^{\text{core}} \,.
\end{equation}
In practice, the localized functions are truncated below a prescribed tolerance to result in sparse matrices $\bPhi_L$ and $\bPhi_L^{\text{core}}$. If they are sufficiently sparse, $\bP^{\phi,\text{core}}$ can be computed in $\order(N)$ complexity.
\vspace{0.1in}
\paragraph{\textbf{Computation of electron-density:}}
One class of commonly employed computational methodologies to reduce the computational complexity of Kohn-Sham DFT calculations constitute the Fermi-operator expansion (FOE) type techniques~\cite{goedecker99,bow2012} which avoid the explicit diagonalization of the discretized Hamiltonian in order to compute the electron-density. The Fermi-operator expansion, which approximates the Fermi-Dirac distribution (cf. equation~\eqref{fermidirac1}) by means of Chebyshev polynomial expansion, is one of the most widely used techniques to realize reduced-order computational complexity in insulating as well as metallic systems. It is important to note that the degree of polynomial required to achieve a desired accuracy in the approximation depends on the spectral width of the discrete Hamiltonian, denoted by $\Delta E$.

In all-electron calculations performed using the finite-element basis, the spectral width can be very large as the largest eigenvalue of the discrete Hamiltonian is $\order (10^6)~Ha$ even for simple systems, owing to the refined finite-element mesh around the nuclei to capture the rapid oscillations in the wavefunctions. Fermi-operator expansion on such large spectral widths can be inaccurate and inefficient. In a recent effort~\cite{motam2014}, this challenge has been addressed by projecting the problem onto a relevant subspace, such as a Chebyshev-filtered space containing the occupied states, and employing the Fermi-operator on the subspace projected Hamiltonian ($\bH^{\phi}$) whose spectral width is commensurate with that of the occupied eigenspectrum. However, one is confronted with another challenge in the case of all-electron calculations, especially those involving heavier atoms. The spectral width of the corresponding Hamiltonian increases as the magnitude of the smallest eigenvalue grows as $\order(Z^2)$ with $Z$ being atomic number and this can deteriorate the accuracy of the Fermi-operator expansion. In the present work, we address this by reformulating the computation of electron-density in terms of the Fermi-Dirac function of the Hamiltonian projected onto the valence-subspace. The spectral width of this valence-subspace projected Hamiltonian is $\order(1Ha)$, irrespective of the materials system, and, thus, Fermi-operator expansion can be employed to develop reduced-order scaling methods for all-electron Kohn-Sham DFT calculations. 

We note that the electron-density is the diagonal of the density matrix and can be written in terms of Fermi-Dirac function of the subspace projected Hamiltonian $f(\bH^{\phi})$ as (cf. equation (60) in Motamarri et al.~\cite{motam2014})
\begin{equation}\label{electrondensity}
 \rho(\bx) =  2\;\overline{\bN}^{T}(\bx)\; \Phi_L \;f(\bH^{\phi}) \;\bS^{-1}\;\Phi_{L}^{T} \;\overline{\bN}(\bx)\,,
 \end{equation}
with
\begin{align*}
\overline{\bN}(\bx) &= \bM^{-1/2}\,\bN(\bx),\,\\ 
\bN(\bx) &= [N_1(\bx)\; N_2(\bx)\; N_3(\bx)\; \cdots\; N_M(\bx)]^{T} \,,
\end{align*}
and
\begin{equation}\label{fermidirac}
f(\bH^{\phi}) = \frac{1}{1 + \exp\left(\frac{\bH^{\phi} - \mu}{\sigma}\right)}
\end{equation}
is the finite-temperature density matrix, with $\mu$ denoting the chemical potential and $\sigma = k_B\,T$. Using equation~\eqref{gammphi}, we can rewrite the above equation in terms of the Fermi-Dirac function of  $\bH^{\phi,\text{val}}$ as
\begin{equation}\label{electrondensity1}
\rho(\bx) =  2\;\overline{\bN}^{T}(\bx)\; \Phi_L \;\left(\bP^{\phi,\text{core}} + f(\bH^{\phi,\text{val}})\right) \;\bS^{-1}\;\Phi_{L}^{T} \overline{\bN}(\bx)\,,
\end{equation}
with $\;\;\bP^{\phi,\text{core}} =\widetilde{\bPhi}_L^{\text{core}}\bS^{\text{core}^{-1}}\widetilde{\bPhi}_L^{\text{core}^{\dagger}}\bS$.

In order to evaluate electron-density using equation ~\eqref{electrondensity1}, we first compute $\bH^{\phi,\text{val}}$, which we recall is given by (cf. equation~\eqref{Hphival})
\begin{equation}
\bH^{\phi,\text{val}} = (\bI - \bP^{\phi,\text{core}})\bH^{\phi}(\bI - \bP^{\phi,\text{core}}) \,.\notag
\end{equation}
We now use Chebyshev polynomial expansion to approximate $f(\bH^{\phi,\text{val}})$, by first scaling and shifting $\bH^{\phi,\text{val}}$ to obtain
$\bH^{\phi,\text{val}}_s$ such that its spectrum lies in $[-1,1]$, and then employing a finite number of Chebyshev polynomials to approximate $f(\bH^{\phi,\text{val}})$ as~\cite{goedecker99,baer97}
\begin{equation}\label{chebexp}
 f(\bH^{\phi,\text{val}}) = \sum_{n=0}^{R} a_n(\sigma_s,\mu_{s}) T_n(\bH^{\phi,\text{val}}_s),
\end{equation}
where
\begin{equation}
\bH^{\phi,\text{val}}_s = \frac{\bH^{\phi,\text{val}} - \bar{\epsilon}}{\Delta \epsilon} \;\;\;;\;\;\sigma_s = \frac{\Delta \epsilon}{\sigma} \;\;;\;\; \mu_{s} = \frac{\mu - \bar{\epsilon} }{\Delta \epsilon},
\end{equation}
\begin{equation}
\Delta \epsilon = \frac{\epsilon^{\phi,\text{val}}_{max} - \epsilon^{\phi,\text{val}}_{min}}{2}\;\;\;\;\;;\;\;\;\;\;\bar{\epsilon} = \frac{\epsilon^{\phi,\text{val}}_{max} + \epsilon^{\phi,\text{val}}_{min}}{2},
 \end{equation}
and
\begin{equation}
  a_n(\sigma_s,\mu_{s})  = \frac{2 - \delta_{n\,0}}{\pi}\int_{-1}^{1} \frac{T_n(x)}{\sqrt{1-x^2}}\,\frac{1}{1 + e^{\sigma_s(x - \mu_{s})}} dx\,,
\end{equation}
where $\delta_{ij}$ denotes the Kronecker delta. In the above, $\epsilon^{\phi,\text{val}}_{max}$ and $\epsilon^{\phi,\text{val}}_{min}$ denote the upper and lower bounds for the spectrum of $\bH^{\phi,\text{val}}$. Estimates for $\epsilon^{\phi,\text{val}}_{max}$ and $\epsilon^{\phi,\text{val}}_{min}$ are computed using the Krylov-Schur method~\cite{kschur2001}. We remark that $\bH^{\phi,\text{val}}$ is a projection onto the valence-subspace represented by localized basis functions, and, hence, $\bH^{\phi,\text{val}}$ is a sparse matrix. Furthermore, if $\bH^{\phi,\text{val}}$ is sufficiently sparse, $f(\bH^{\phi,\text{val}})$ can be computed in $\order(N)$ complexity~\cite{baer97}. Most importantly, the degree $R$ of the Chebyshev expansion in ~\eqref{chebexp} is proportional to the spectral width $\Delta E = \epsilon^{\phi,\text{val}}_{max}- \epsilon^{\phi,\text{val}}_{min}$ of $\bH^{\phi,\text{val}}$, which is $\order (1\,Ha)$. We remark that this spectral width is much smaller than that of $\bH^{\phi}$, thus, allowing an accurate and efficient computation of Fermi-operator expansion using a Chebyshev polynomial expansion of $\order (100)$ for all-electron calculations on any materials system as subsequently demonstrated in the numerical studies.

The Fermi-energy ($\mu$), which is required in the computation of the Fermi-operator expansion of $f(\bH^{\phi,\text{val}})$, is evaluated using the constraint given in ~\eqref{valconst}
\begin{equation}
2\;\text{tr}\left(f(\bH^{\phi,\text{val}})\right) = N_{e,\text{val}} = N_e - 2\,N_{\text{core}} \,.\notag
\end{equation}
Finally, the band energy ($E_b$), which can also be expressed in terms of $\bH^{\phi,\text{val}}$, is evaluated as
\begin{equation}
E_b = 2\,\text{tr}\left(\left[\widetilde{\bPhi}_L^{\text{core}}\bS^{\text{core}^{-1}}\widetilde{\bPhi}_L^{\text{core}^{\dagger}}\bS + f(\bH^{\phi,\text{val}})\right]\bH^{\phi}\right)\,.
\end{equation}

\section{Results and discussion}\label{sec:results}
The accuracy and performance of the proposed spectrum-splitting method is investigated in this section. The benchmark systems considered in this study involve non-periodic three dimensional materials systems with moderate to high atomic numbers. These include silicon nano-clusters of varying sizes, containing $1\times 1\times 1$ (252 electrons), $2\times 1\times 1$ (434 electrons), $2 \times 2 \times 2$ (1330 electrons) diamond-cubic unit-cells, as well as a single gold atom and a six-atom gold nano-cluster. 

In order to assess the accuracy of the proposed approach, we use as reference the ground-state energy of the Kohn-Sham problem solved using the Chebyshev-filtered subspace iteration method for the finite-element basis~\cite{motam2013} (ChFSI-FE). The ChFSI-FE involves the projection of the discrete Hamiltonian onto the Chebyshev-filtered subspace spanned by an orthogonal basis, followed by an explicit computation of the eigenvalues and eigenvectors via diagonalization of the projected Hamiltonian to estimate the electron-density. In all our simulations---reference calculations, as well as, the benchmark calculations discussed subsequently---we use the n-stage Anderson~\cite{andmix65} mixing scheme on the electron-density in self-consistent field iteration of the Kohn-Sham problem with a stopping criterion of $10^{-8}$ in the square of the $L^2$ norm of the change in electron-density in two successive iterations. 

In order to demonstrate the effectiveness of the proposed spectrum-splitting method, we compare its performance with the subspace projection algorithm (SubPJ-FE) proposed in Motamarri et.al~\cite{motam2014}. In SubPJ-FE the Fermi-operator expansion is employed on the Chebyshev-filtered subspace projected Hamiltonian as opposed to the valence-subspace projected Hamiltonian in the proposed spectrum-splitting method, with all else being identical in both sets of calculations. The ground-state energies for each of these benchmark systems are computed for varying polynomial degrees $R$ used in the Chebyshev expansion of the Fermi-Dirac function of the projected Hamlitonians, $\bH^{\phi,\text{val}}$ in the proposed spectrum-splitting approach and $\bH^{\phi}$ in SubPJ-FE. The error in the ground-state energies obtained in each of the two methods is measured with respect to the reference energy obtained using ChFSI-FE and is plotted against $R$. 

\subsection{Silicon}
We consider silicon nano-clusters containing $1\times 1\times 1$ (252 electrons), $2\times 1\times 1$ (434 electrons), $2 \times 2 \times 2$ (1330 electrons) diamond-cubic unit cells with a lattice constant of $10.26~a.u.$, and conduct all-electron calculations to test the performance of the spectrum-splitting approach. Finite-element meshes with fifth-order spectral finite-elements (HEX216SPECT) are chosen such that the discretization error is less than $5~mHa$ in the ground-state energy per atom. Identical finite-element meshes are employed in the benchmark calculations presented below as well as the reference calculations. 

\begin{figure}[htbp]
\begin{center}
\includegraphics[width=0.45\textwidth]{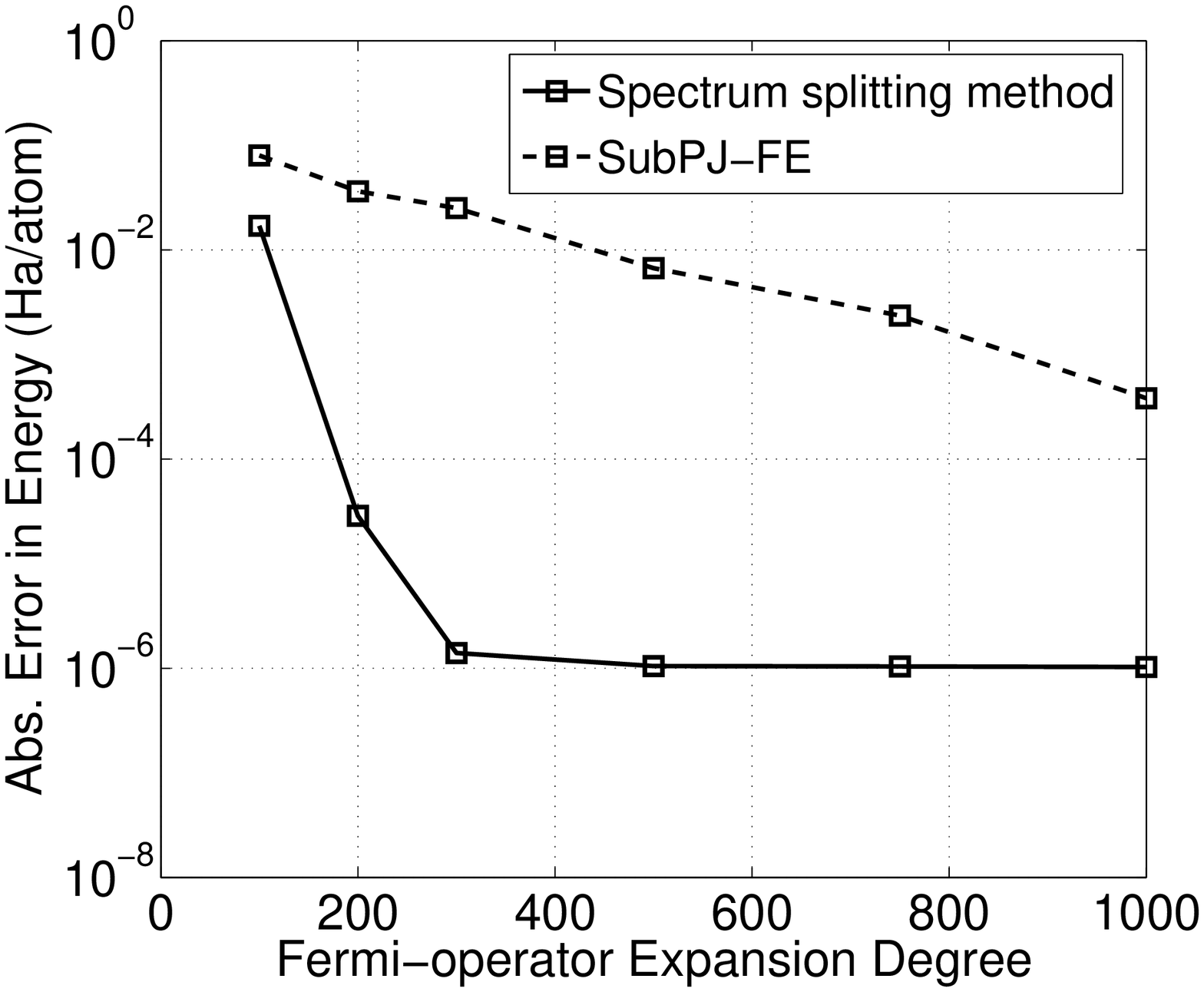}
\caption{\small{Comparison between the spectrum-splitting method and SubPJ-FE. Case study: Silicon 1x1x1 nano-cluster.}}
\label{fig:silicon18}
\end{center}
\end{figure}

\begin{figure}[htbp]
\begin{center}
\includegraphics[width=0.45\textwidth]{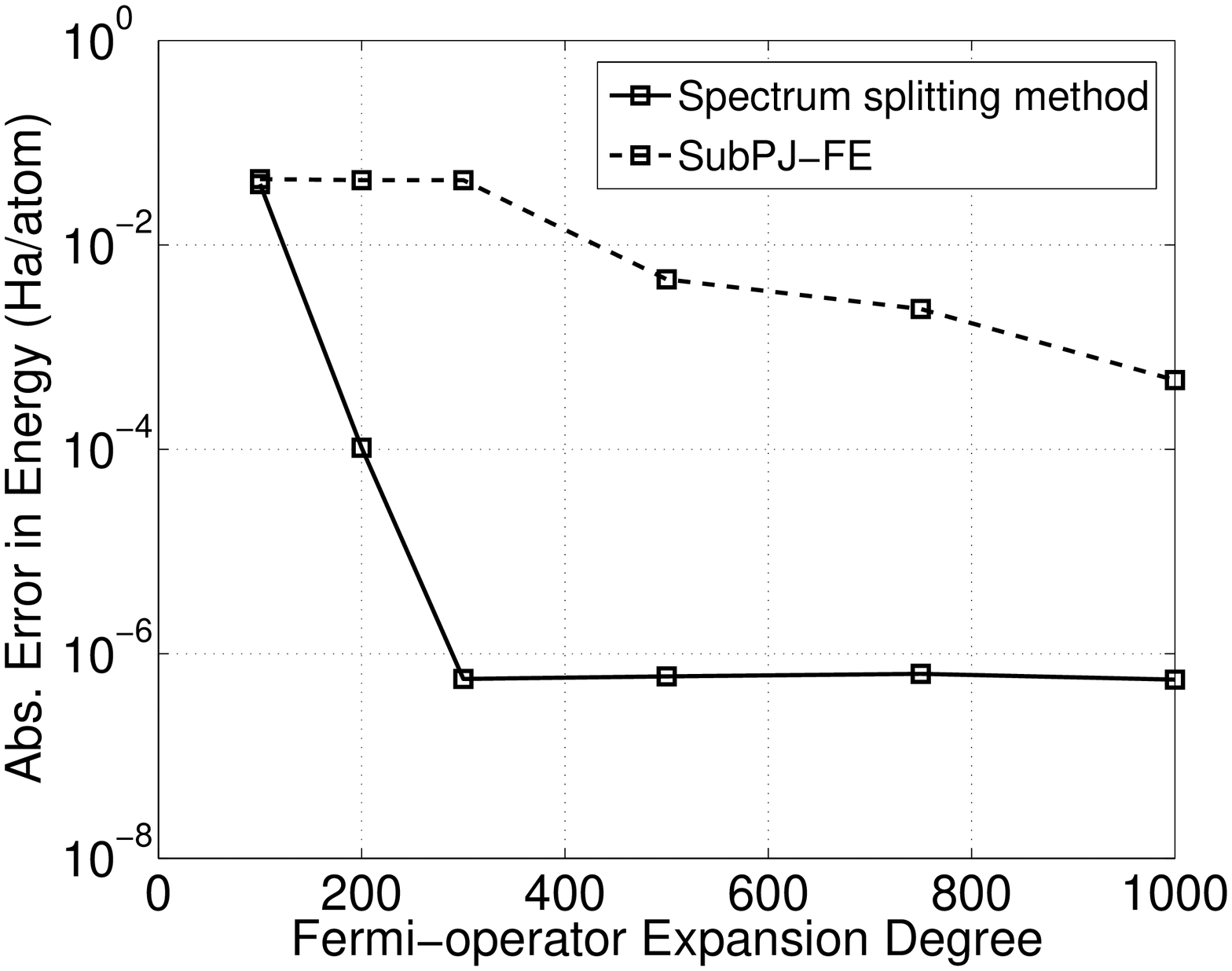}
\caption{\small{Comparison between the spectrum-splitting method and SubPJ-FE. Case study: Silicon 2x1x1 nano-cluster.}}
\label{fig:silicon31}
\end{center}
\end{figure}

Figures~\ref{fig:silicon18} and ~\ref{fig:silicon31} show the comparison between the proposed spectrum-splitting method and SubPJ-FE for silicon nano-clusters containing $1\times 1\times 1$ (252 electrons), $2\times 1\times 1$ (434 electrons) unit-cells respectively. A Fermi-smearing parameter of $0.003262~Ha$ (T=1000K) is employed in these simulations. 
These results indicate that the proposed method provides significantly better accuracies in the ground-state energies in comparison to SubPJ-FE for similar polynomial degrees. Accuracies in the ground-state energies of better than $10^{-4}~Ha/atom$ with reference calculations can be obtained with just a polynomial degree $R$ of $200$ using the spectrum-splitting method, whereas $R$ greater than 1000 is required to get to accuracies close to $10^{-4}~Ha/atom$ with SubPJ-FE. While the error in the ground-state energy decreases exponentially with increasing $R$ in the spectrum-splitting method, the error is seen to stagnate at around $10^{-6}~Ha/atom$. This is due to competing non-variational errors arising from numerical integration involved in the evaluation of coefficients in the Chebyshev polynomial expansion~\eqref{chebexp}. 

\begin{figure}[htbp]
\begin{center}
\includegraphics[width=0.5\textwidth]{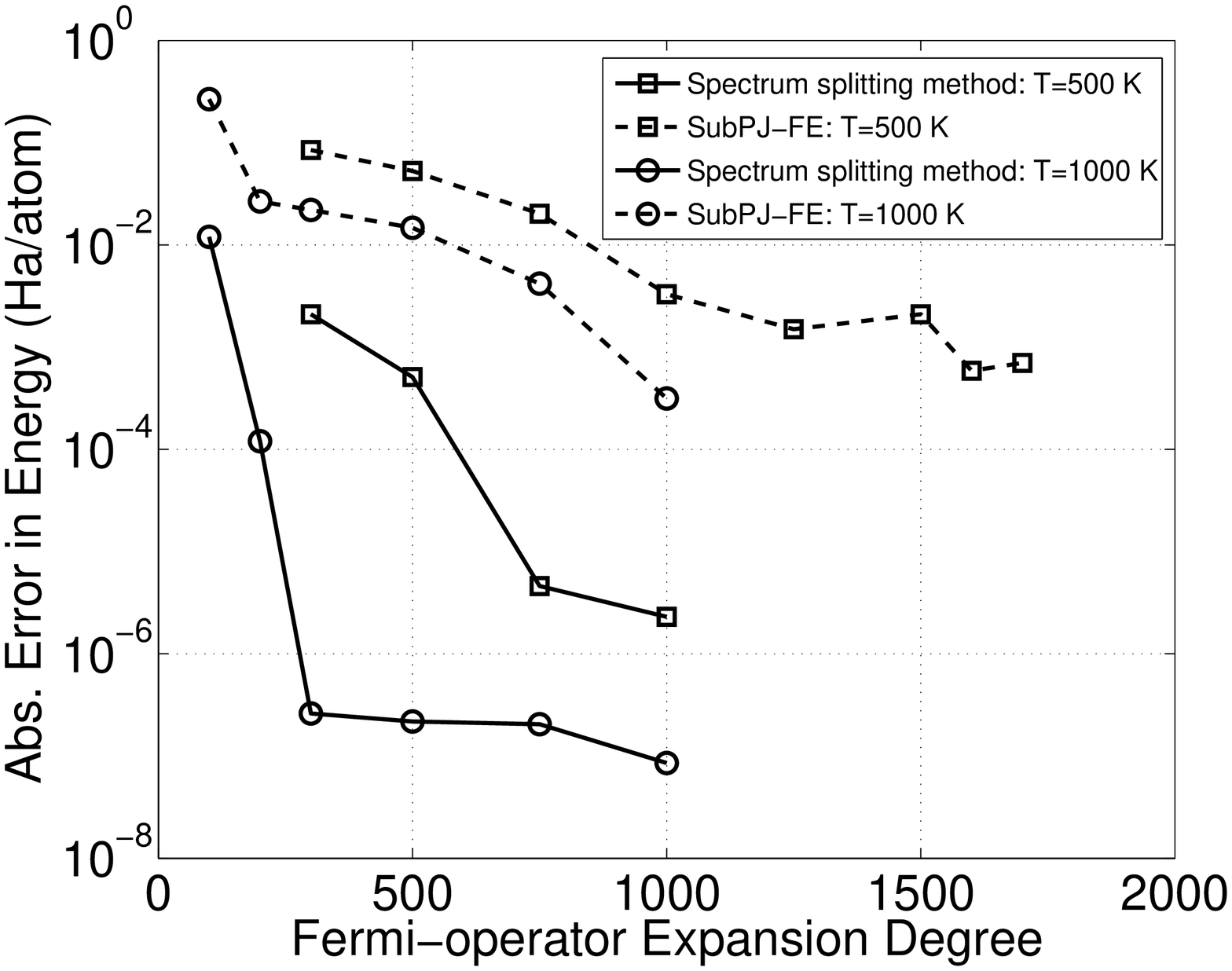}
\caption{\small{Comparison between the spectrum-splitting method and SubPJ-FE. Case study: Silicon 2x2x2 nano-cluster.}}
\label{fig:silicon95}
\end{center}
\end{figure}

We next consider the benchmark calculations on silicon nano-cluster containing $2 \times 2 \times 2$ (1330 electrons) unit-cells for the case of two smearing parameters: $0.00163~Ha$(T=500K) and $0.003262~Ha$ (T=1000K). Figure~\ref{fig:silicon95} shows the comparison between the proposed method and SubPJ-FE for this benchmark system.
The results from Figure~\ref{fig:silicon95} demonstrate that the accuracy and effectiveness of the proposed method is retained for larger materials systems. We note a five-fold reduction in the polynomial degree for accuracies of $10^{-4}~Ha/atom$ with a temperature of 1000K for the spectrum-splitting approach in comparison to SubPJ-FE---a similar reduction as observed in the smaller-sized nano-clusters. Furthermore, the simulations with a lower temperature of 500K show that $R\sim600$ results in accuracies better than $10^{-5}~Ha/atom$  with the proposed approach, whereas even with $R\sim 1500$ we observe errors of $\sim 10^{-3}~Ha/atom$ with SubPJ-FE.

\subsection{Gold}
In order to further investigate the performance of the proposed method for material systems with larger atomic numbers, we consider gold (Atomic number 79) as our model example. We consider two benchmark problems: (i) single atom gold (Au); (ii) a planar gold cluster~\cite{goldcluster}  containing six atoms (Au$_6$) with Au-Au bond length of 5.055 a.u. We choose finite-element meshes with fifth-order spectral finite-elements (HEX216SPECT) for these benchmark examples such that the discretization error is less than $5~mHa$.

\begin{figure}[htbp]
\begin{center}
\includegraphics[width=0.45\textwidth]{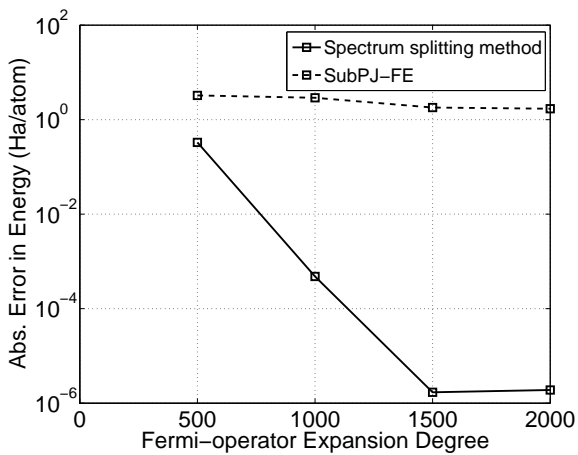}
\caption{\small{Comparison between the spectrum-splitting method and SubPJ-FE. Case study: Au single atom.}}
\label{fig:gold}
\end{center}
\end{figure}

\begin{figure}[htbp]
\begin{center}
\includegraphics[width=0.45\textwidth]{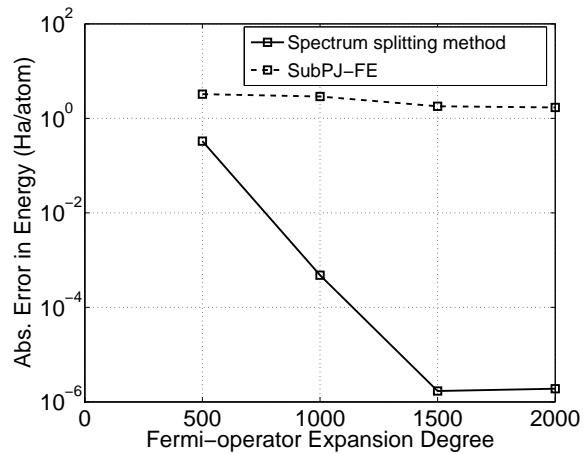}
\caption{\small{Comparison between the spectrum-splitting method and SubPJ-FE. Case study: Au$_6$ nano-cluster.}}
\label{fig:gold6}
\end{center}
\end{figure}

The error in the ground-state energies, measured with respect to the reference energies obtained using ChFSI-FE, for the proposed spectrum-splitting method and SubPJ-FE are computed for various polynomial degrees in the Chebyshev expansion of the Fermi-operator. Figures ~\ref{fig:gold} and ~\ref{fig:gold6} show these results for case of Au single atom  and Au$_6$ nano-cluster, respectively. A Fermi-smearing parameter of $0.00163~Ha$ (T=500K) is used in these calculations. 

These results demonstrate that the accuracy of the proposed method is far superior than SubPJ-FE. In particular, accuracies in the ground-state energies close to $10^{-4}~Ha/atom$ can be obtained with a polynomial degree that is a little over $1000$ using the proposed spectrum-splitting method, whereas SubPJ-FE resulted in ground-state energies with errors close to $\order(1Ha)$ per atom even with $R$ greater than 2000. These benchmark studies on Au underscores the advantage of the proposed spectrum-splitting approach in employing Fermi-operator type techniques for material systems with heavy atomic numbers.

\section{Summary}\label{sec:summary}
In the present work, we formulated a spectrum-splitting approach for employing Fermi-operator expansion in all-electron Kohn-Sham DFT calculations, and presented the approach in the framework of spectral finite-element discretization of the Kohn-Sham DFT problem. In the proposed method, in every iteration of the self-consistent field procedure, an eigen-subspace containing the occupied states of the Kohn-Sham Hamiltonian is computed using a Chebyshev filter. The Kohn-Sham Hamiltonian is projected onto the Chebyshev-filtered subspace spanned by localized basis functions which are computed using a localization procedure. The core-subspace in the Chebyshev-filtered subspace is computed using another appropriately constructed Chebyshev filter, and the valence-subspace is extracted as its complement. The Fermi-operator expansion of the Kohn-Sham Hamiltonian is subsequently evaluated as the sum of the projection operator corresponding to the core-subspace and the Fermi-operator expansion of the valence-subspace projected Kohn-Sham Hamiltonian. As the spectral width of the valence-subspace projected Kohn-Sham Hamiltonian is $\order(1\,Ha)$, and is independent of the materials system and discretization, the Fermi-operator expansion can be employed on all-electron Kohn-Sham DFT calculations.      

The accuracy and performance of the proposed method was investigated on two different materials systems: (i) Silicon with a moderate atomic number, and (ii) Gold with a high atomic number. The benchmark systems involved silicon nano-clusters up to 1330 electrons, a single gold atom, and a six-atom gold cluster. In all the cases, the proposed spectrum-splitting method provided ground-state energies that are in excellent agreement with reference calculations. In particular, in the case of silicon nano-clusters, the proposed spectrum-splitting approach resulted in a five-fold reduction in the Fermi-operator expansion polynomial degree to achieve accuracies close to $10^{-4} Ha/atom$ in the ground-state energies. Further, the effectiveness of the proposed approach was even more significant in the case of gold, where the spectrum-splitting approach was found to be indispensable in achieving chemical accuracy.

Fermi-operator expansion type techniques avoid explicit diagonalization of the Kohn-Sham Hamiltonian, and offer a viable path to developing reduced-order scaling methods to solve the Kohn-Sham problem for both insulating and metallic systems. The proposed spectrum-splitting approach extends the applicability of Fermi-operator type expansion methods to all-electron DFT calculations, independent of the materials system and discretization. The proposed spectrum-splitting approach in conjunction with enriched finite-element basis, where the finite-element basis is enriched with compactly supported single-atom Kohn-Sham orbitals, has the potential to enable all-electron DFT calculations on system sizes not accessible heretofore.

\begin{acknowledgments}
We gratefully acknowledge the support from the Air Force Office of Scientific Research under Grant No. FA9550-13-1-0113, and the support from the U.S. Army Research Laboratory (ARL) through the Materials in Extreme Dynamic Environments (MEDE) Collaborative Research Alliance (CRA) under Award Number W911NF-11-R-0001. V.G. also acknowledges the hospitality of the Division of Engineering and Applied Sciences at the California Institute of Technology while pursuing this work. We also acknowledge Advanced Research Computing at University of Michigan  through the Flux computing platform, and Extreme Science and Engineering Discovery Environment (XSEDE), which is supported by National Science Foundation grant number ACI-1053575, for providing the computing resources.
\end{acknowledgments}

\end{document}